\documentclass[amsmath,amssymb,prb,superscriptaddress,twocolumn,showpacs,longbibliography,floatfix]{revtex4-1}
\usepackage{graphicx}
\usepackage{dcolumn}
\usepackage[colorlinks=true,linkcolor=blue,citecolor=blue,urlcolor=blue]{hyperref}
\usepackage{multirow}
\usepackage[usenames,dvipsnames]{xcolor}
\usepackage{soul}
\usepackage{braket}
\usepackage{bm}
\usepackage{nicefrac}
\usepackage{siunitx}
\usepackage{color}

\newcommand{\HH}{\mathcal{H}}

\newcommand{\NN}{\mathcal{N}}
\newcommand{\ud}{\mathrm{d}}
\newcommand{\iu}{\mathrm{i}}
\newcommand{\Tr}{\mathrm{Tr}}

\newcommand{\RE}{\mathrm{Re}}

\newcommand{\VEC}[1]{\mathbf{#1}}

\begin{document}

\title{First-principles investigation of spin-wave dispersions in surface-reconstructed Co thin films on W(110)}

\author{Flaviano Jos\'e dos Santos}\email{f.dos.santos@fz-juelich.de}
\author{Manuel dos Santos Dias}
\author{Samir Lounis}\email{s.lounis@fz-juelich.de}
\affiliation{Peter Gr\"{u}nberg Institut and Institute for Advanced Simulation, Forschungszentrum J\"{u}lich \& JARA, D-52425 J\"{u}lich, Germany}

\date{\today}

\begin{abstract}
\noindent
We computed spin-wave dispersions of surface-reconstructed Co films on the W(110) surface in the adiabatic approximation.
The magnetic exchange interactions are obtained via first-principles electronic-structure calculations using the Korringa-Kohn-Rostoker Green-function method.
We analyze the strength and oscillatory behavior of the intralayer and interlayer magnetic interactions and investigate the resulting spin-wave dispersions as a function of the thickness of Co films.
In particular, we highlight and explain the strong impact of hybridization of the electronic states at the Co--W interface on the magnetic exchange interactions and on the spin-wave dispersions.
We compare our results to recent measurements based on electron-energy-loss spectroscopy [E.~Michel, H.~Ibach, and C.M.~Schneider, Phys. Rev. B \textbf{92}, 024407 (2015)].
Good overall agreement with experimental findings can be obtained by considering the possible overestimation of the spin splitting, stemming from the local-spin-density approximation, and adopting an appropriate correction.
\end{abstract}

\pacs{}

\keywords{spin-waves, thin films, cobalt on tungsten}

\maketitle

\section{Introduction}\label{sec:intro}

Novel ways to transmit, process, and store information in solid-state devices have become increasingly important, as improvement in current technology achieved solely via miniaturization is reaching its physical limits.
Spintronics is a prime candidate, where the spin of the electrons is exploited to perform the basic device processes~\cite{Zutic2004,Maekawa2012}.
One of the manifestations of electron spin are spin-waves, which are collective excitations of magnetic systems.
The excitation quantum is called magnon, corresponding to a net lowering of spin angular momentum by $\hbar$.
The application of spin-waves to transmit and process information in novel structures and devices defines the field of magnon spintronics~\cite{Chumak2015}.
For example, spin-waves can be used to move efficiently domain walls in race-track memories and logic gates; could allow for information transmission over macroscopic distances free of Ohmic losses; can be applied in a wide operational frequency range, which is suitable for various applications, e.g., telecommunication systems and computing; or can even make possible the realization of wave-based computing~\cite{Chumak2015}.

Magnetism in thin films features special properties driven by the low dimensionality of these systems, such as atoms presenting enhanced magnetic moments with respect to their bulk values, which approach the atomic limit, and novel non-trivial magnetic phases such as spin spiral and skyrmions~\cite{Asada1999,Gradmann1993,Vaz2008,Nagaosa2013,Wiesendanger2009}.
This, combined with the prospect of creating smaller devices, has directed great attention to these nanostructures.
Many studies are focusing on the dynamical magnetic properties (e.g.,~spin-waves), as key properties such as the time it takes to switch a magnetic bit or to transport magnetic information are dynamical processes.
A technique that is very suitable to study spin-waves in thin films is electron-energy-loss spectroscopy (EELS) or its spin-polarized version (SPEELS)~\cite{Etzkorn2007,Mills1967,Ibach2014,Plihal1999,Vollmer2003}.
Electrons have a much larger scattering cross section than neutrons (routinely used to probe spin-waves in bulk materials), which compensates for the much smaller scattering volume of a thin magnetic film.
Recently, another kind of inelastic electron spectroscopy was shown to be able to detect spin-wave modes, utilizing scanning tunneling microscopy~\cite{Balashov2014}.
This can be seen as a real-space complementary technique to the reciprocal space picture afforded by (SP)EELS.

EELS consists in shooting an electron beam with well-defined energy at a target magnetic surface.
The scattering geometry determines the in-plane momentum transfer, and the energy of the scattered electrons is determined by a detector~\cite{Ibach2014}.
Among other degrees of freedom, the transferred energy and momentum can be converted into the creation or destruction of magnons.
The relation between a magnon's energy and its momentum is called the magnon (or spin-wave) dispersion relation (dispersion for short), and it will be a key quantity through out this work.
The study of spin-waves at surfaces and thin films by EELS was already proposed back in 1967 by Mills~\cite{Mills1967}.
However, it took more than three decades until specialized electron spectrometers were built, allowing the first investigations~\cite{Plihal1999,Vollmer2003}.

Originally only one spin-wave mode was experimentally observed~\cite{Vollmer2003}.
This was in stark contrast with the theoretical expectation of one mode per layer of a uniform ferromagnetic thin film, based on a simple Heisenberg model.
A more sophisticated theoretical description, taking into account the spin-wave suppression due to Landau damping (decay into Stoner excitations), also predicted that more modes should be observed~\cite{Costa2004}.
This has been recently borne out experimentally, due to a large improvement in the energy resolution of the EELS spectrometers (now $\sim 2$ meV), making it possible for the first time to resolve up to three spin-wave modes~\cite{Michel2015,Michel2016_1,Michel2016_2}.
Faced with such a wealth of experimental results, it becomes essential to perform detailed theoretical investigations, in order to ascertain the quality of the current methods and our understanding of the underlying physics. 

Consequently, this paper concerns the theoretical properties of spin-waves in thin films of cobalt deposited on tungsten (110), following the work of Michel \textit{et al.}~\cite{Michel2015}, which revisits the initial investigation of Vollmer 
\textit{et al.}~\cite{Vollmer2003} This system is peculiar since a realistic simulation of its electronic properties requires us to consider the surface reconstruction of Co thin films, leading to dramatic computational costs because of the large supercells with several inequivalent atoms.
We perform first-principles electronic structure calculations, extracting the magnetic exchange interactions and computing the spin-wave dispersion in the adiabatic approximation~\cite{Halilov1998,Antropov1995}.
This approach has some limitations, stemming from the neglect of the interaction between the collective spin-wave modes and the continuum of Stoner excitations.
This leads to Landau damping~\cite{Cooke1980,Muniz2002}, which can heavily damp the spin-wave modes, and it may also renormalize the spin-wave energies.
However, it has been argued theoretically and demonstrated by explicit calculations~\cite{Costa2004,Costa2004_2,Buczek2010} that the Heisenberg model description is reasonable for low spin-wave energies and not too large wavevectors, which is precisely the range relevant to (SP)EELS and the one in which we are interested.

Although it is responsible for the magnetocrystalline anisotropy (MCA) and for the Dzyaloshinskii-Moriya interaction (DMI), we left out spin-orbit coupling from the calculations.
The MCA determines the ferromagnetic easy axis and leads to the zero wavevector gap in the spin-wave dispersion, while the DMI favors the formation of noncollinear magnetic structures.
In Ref.~\onlinecite{Michel2015} the MCA gap is unresolvable and no DMI-induced asymmetries in the spin-wave dispersions were measured, which must be due to the strong ferromagnetic exchange interactions.
In view of the considerable complexity of the problem already without relativistic effects, we chose to leave this aspect for future investigations.

Our paper is organized as follows.
We first summarize our theoretical approach in Sec.~\ref{sec:theory}.
Then the ground-state properties obtained from first-principles calculations are analyzed in Sec.~\ref{sec:groundstate}, including the magnetic exchange interactions.
The latter are then used in Sec.~\ref{sec:swdispersion} to compute the adiabatic spin-wave dispersions.
Finally, our conclusions are gathered in Sec.~\ref{sec:conc}.

\vspace{-2.5em}

\section{Methods}\label{sec:theory}

Co thin films deposited on W(110) are inhomogenous ferromagnets.
Besides the vertical inhomogeneity due to the layered structure, there is also lateral inhomogeneity, due to a surface reconstruction.
In this section, we summarize the linear spin-wave theory for inhomogeneous ferromagnets, and we explain why comparison with experimental results requires an unfolding of the computed spin-wave band structure.

\subsection{Spin-waves in an inhomogeneous ferromagnet}

The magnetic moments $\VEC{M}_{i\mu}$ are taken as classical vectors of constant length, $\VEC{M}_{i\mu} = M_{\mu} \VEC{m}_{i\mu}$.
In terms of the unit vectors $\VEC{m}_{i\mu}$, the Heisenberg model for an inhomogeneous ferromagnet can be written as
\begin{equation}
  \label{eq:heisenberghamil}
  \HH = -\frac{1}{2}\sum_{i\mu}\sum_{j\nu} J_{i\mu,j\nu}\,\VEC{m}_{i\mu}\cdot\VEC{m}_{j\nu} \quad .
\end{equation}
Here $i,j$ label unit cells forming a Bravais lattice, while $\mu,\nu$ run over the $N_b$ basis atoms.
The position of a particular magnetic moment is given by $\VEC{R}_{i\mu} = \VEC{R}_i + \VEC{R}_\mu$.
The magnetic exchange interactions are symmetric, $J_{j\nu,i\mu} = J_{i\mu,j\nu}$, and depend only on the distance between unit cells, $J_{i\mu,j\nu} = J_{\mu\nu}(\VEC{R}_i-\VEC{R}_j)$.

The adiabatic spin dynamics are described by the Landau-Lifshitz equation of motion:
\begin{equation}
  \label{eq:llequation}
  \frac{M_\mu}{\gamma}\,\frac{\ud\VEC{m}_{i\mu}}{\ud t} = -\VEC{m}_{i\mu} \times \VEC{B}_{i\mu}^{\text{eff}} \quad,
\end{equation}
with the gyromagnetic ratio $\gamma = 2$ (spin-only) and the effective magnetic field
\begin{equation}
  \VEC{B}_{i\mu}^{\text{eff}} = -\frac{\partial\HH}{\partial\VEC{m}_{i\mu}} = \sum_{j\nu} J_{i\mu,j\nu}\,\VEC{m}_{j\nu} \quad .
\end{equation}

Consider the spin-wave ansatz in the small-amplitude limit ($\theta_\mu \ll 1$):
\begin{subequations}
  \label{eq:linearsw}
  \begin{equation}
    \VEC{m}_{i\mu}(\VEC{q},t) \approx \VEC{n}^z + \theta_\mu\,\VEC{n}_{i\mu}^\perp(\VEC{q},t) + \mathcal{O}\big(\theta_\mu^2\big) \quad ,
  \end{equation}\vspace{-1em}
  \begin{equation}
    \VEC{n}_{i\mu}^\perp (\VEC{q},t) = \RE\left[e^{\iu(\VEC{q}\cdot\VEC{R}_i + \phi_\mu - \omega t)}\left(\VEC{n}^x + \iu\,\VEC{n}^y\right)\right] \quad .
  \end{equation}
\end{subequations}
The unit vector in the $\alpha$-direction is denoted $\VEC{n}^\alpha$.
The magnetic moments are assumed to point in the $z$-direction in the ground state, with $\theta_\mu$ the cone angle of the spin-wave precession.

Inserting our ansatz, Eqs.~\eqref{eq:linearsw}, in the equation of motion, Eq.~\eqref{eq:llequation}, leads to the eigenvalue problem:~\cite{Halilov1998}
\begin{equation}
  \label{eq:linearll}
  \sum_\nu J_{\mu\nu}(\VEC{q})\,u_\nu - \sum_\nu J_{\mu\nu}(\VEC{0})\,u_\mu = \frac{M_\mu}{\gamma}\,\omega\,u_\mu \quad ,
\end{equation}
with $u_\mu = \theta_\mu\,e^{\iu\phi_\mu}$ being the eigenvectors and the lattice Fourier transform of the exchange interactions,
\begin{equation}
  J_{\mu\nu}(\VEC{q}) = \sum_{j} e^{\iu\VEC{q}\cdot(\VEC{R}_j-\VEC{R}_i)}\,J_{i\mu,j\nu} \quad .
\end{equation}
The substitution $u_\mu = \sqrt{\gamma/M_\mu}\,\tilde{u}_\mu$ leads to a standard eigenvalue problem.
For every wavevector $\VEC{q}$ there are $N_b$ spin-wave branches, with frequency $\omega_n(\VEC{q}) \geq 0$ and eigenvector $\tilde{u}_\mu^n(\VEC{q})$.

The input quantities for the spin-wave calculations (the magnetic moments, $M_\mu$, and the magnetic exchange interactions, $J_{i\mu,j\nu}$) can be obtained from first-principles.
The eigenvalues and eigenvectors obtained by solving {Eq.~\eqref{eq:linearll}} can be summarized in a single quantity, the spectral density matrix,
\begin{align}\label{eq:Glmulnu}
  \rho_{\mu\nu}(\VEC{q},\omega) &= \sum_{n=1}^{N_b} \delta\big(\omega - \omega_n(\VEC{q})\big)\,\tilde{u}_{\mu}^n(\VEC{q}) \left(\tilde{u}_{\nu}^n(\VEC{q})\right)^* \quad .
\end{align}
This quantity is at the heart of the unfolding method presented in the next section.
The practical use of Eq.~\eqref{eq:Glmulnu} requires a numerical representation of the delta function, for which we employ the Lorentzian function $\delta(\omega)\simeq (\eta/\pi)/(\omega^2+ \eta^2)$, introducing the broadening parameter $\eta$.

\subsection{Link to inelastic scattering experiments}

The intrinsic spin-wave spectrum is not necessarily simply related to the experimental measurements.
One must consider how the experiment probes the spin-wave properties of the system.
As we explained in the introduction, newly developed high energy resolution EELS is one of the motivations for this work.
A complete description of an EELS experiment requires a detailed multiple scattering analysis of the probing electrons, taking into account all inelastic effects.
As such a description is highly involved, here we make some considerations aimed at justifying a simpler connection between theory and experiment.
We assume that the differential inelastic scattering cross section between an initial probe state (with energy $E_i$, momentum $\VEC{k}_i$ and spin $s_i$) and final probe states (with $E_{\!f}$, $\VEC{k}_{\!f}$ and $s_{\!f}$), which are collected in a solid angle window $\ud\Omega$, is proportional to the dynamic structure factor:~\cite{Marder2010}
\begin{equation}
  \frac{\ud\sigma}{\ud\omega\,\ud\Omega} \propto \mathcal{S}^{s_{\!f}s_i}(\bm{q},\omega) \qquad ,
\end{equation}
with the energy and momentum transfer defined as
\begin{equation}
  E_i - E_{\!f} = \omega \;\;,\quad \VEC{k}_i - \VEC{k}_{\!f} = \VEC{q} + \VEC{G}
\end{equation}
$\omega > 0$ corresponds to energy absorption and vice-versa.
Momentum is conserved up to a reciprocal lattice vector, $\VEC{G}$, except for non-periodic directions (for films only in-plane momentum is fixed by Bragg scattering).

For inelastic scattering involving spin-waves (relatively low energy compared to that of the probing electron beam), it is usually a fair approximation to assume that the dynamic structure factor is proportional to the density of spin-wave excitations $\NN_{l'l}(\VEC{q},\omega)$ (here at zero temperature),~\cite{Gokhale1992}
\begin{align}
  \label{eq:sfactor}
  \mathcal{S}(\VEC{q},\omega) \propto \sum_{l'l} A_{l'l} \sqrt{M_{l'} M_l}\,\NN_{l'l}(\VEC{q},\omega) \quad.
\end{align}
With the application to layered systems in mind, we have split the basis index into two, $\VEC{R}_{\mu} \rightarrow \VEC{R}_{l\mu} = \VEC{R}_l + \VEC{b}_\mu$, with $\VEC{R}_l$ the origin for layer $l$, and $\VEC{b}_\mu$ the location of a basis atom with respect to the origin of layer $l$.
The spectral density matrix is now given by
\begin{align}
  \NN_{l'l}(\VEC{q},\omega) = \frac{1}{N_b} \sum_{\mu\nu} e^{\iu \VEC{q}\cdot \VEC{b}_{\mu\nu} } 
  \rho_{l'\mu,l\nu}(\VEC{q},\omega) \quad ,\label{eq:Nll}
\end{align}
with the vector $\VEC{b}_{\mu\nu} = \VEC{b}_\nu - \VEC{b}_\mu$.
In a scattering experiment, due to the wave nature of the probing beam, the response of every atom arrives to the detector with different phases.
The resulting interference is destructive for most modes arising from atoms which are crystallographically nearly equivalent (for example in the same layer); the waves that interfere constructively lead to the experimentally detected signal.
Such phase differences are encoded in the Fourier factor of Eq.~\eqref{eq:Nll}, and define the unfolding of the computed spin-wave bands.

The factor $A_{l'l}$ describes how the experimental probes (electrons, for instance) couple to the intrinsic spin-wave excitations, and has assumed many forms in the literature.
Taroni \textit{et al.}~{\cite{Taroni2011}} consider $A_{l'l}=e^{\iu \VEC{q}\cdot \VEC{R}_{l'l}}$, with $\VEC{R}_{l'l} = \VEC{R}_{l} - \VEC{R}_{l'}$, and they show that this choice leads to the suppression of the optical spin-wave modes in $\mathcal{S}(\VEC{q},\omega)$.
With this particular choice, the probed system is excited uniformly which leads to the acoustic mode only.
Using arguments from scattering theory, Rajeswari \textit{et al.}~{\cite{Rajeswari2012}} proposed $A_{l'l} = e^{-(z_{l'}+z_l)/\lambda_{\text{d}}}\,e^{\iu \VEC{q}\cdot \VEC{R}_{l'l}}$, where $z_l$ is the distance between layer $l$ and the surface of the film, and $\lambda_{\text{d}}$ is the finite penetration depth of the electron beam. 
This explains the experimental detection of optical modes in the EELS experiment.

\begin{figure}[b]
  \centering
  \includegraphics[scale=1.2,clip=true,trim=0.0em 0em 0em 0em]{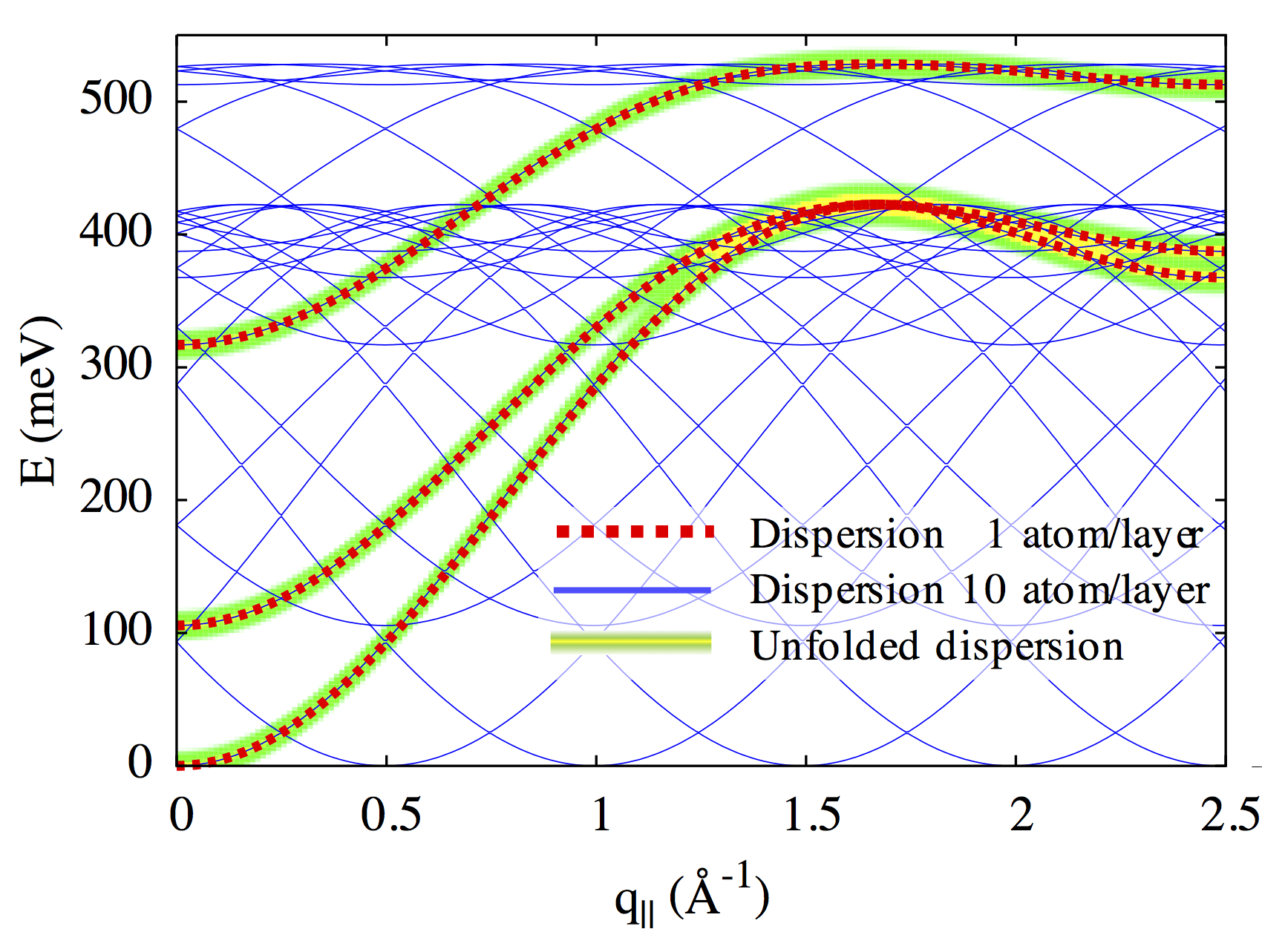} \vspace{-2em}
  \caption{\label{fig:unfolding_example} Illustrating the unfolding scheme.
  The spin-wave dispersion of a uniform trilayer  is calculated using {Eq.~\eqref{eq:linearll}} with 1 and 10 atoms per layer, red dashed lines and blue solid lines respectively.
  We considered nearest-neighbour intralayer and interlayer magnetic interactions $J = 9\,\text{meV}$ and a moment of $1\,\mu_{\text{B}}$ for all atoms.
  The unfolding (green-yellow color map) is obtained via {Eq.~\eqref{eq:sfactor}}, with our choice of $A_{l'l} = \delta_{l'l}$.
  }
\end{figure}

In this work, we are solely interested in the spin-wave dispersion relation, rather than their spectral line shapes and intensities, as they cannot be accessed within the frozen magnon approximation since electron-hole excitations are not included.
Therefore, we introduce $A_{l'l}=\delta_{l'l}$, which gives equal weight to the contributions of each layer to the intensity of a given spin-wave mode, and it yields the layer-resolved density of spin-wave excitations.
This is an appropriate choice to trace out the dispersion of each spin-wave branch throughout the entire Brillouin zone without a parameter-dependent intensity function.
We emphasize that none of the choices mentioned above for $A_{l'l}$ affects the spin-waves energy dispersion, but only the intensities of the bands.
The unfolding procedure is illustrated in {Fig.~\ref{fig:unfolding_example}} for the trivial case of a trilayer with uniform nearest-neighbour magnetic interaction, described in two ways: with a basis of one atom per layer and with a basis of ten atoms per layer.
The spin-wave dispersions computed from {Eq.~\eqref{eq:linearll}} then comprise 3 and 30 branches, respectively, as can be seen in the figure.
Applying {Eq.~\eqref{eq:sfactor}} to the case of the 30 bands shows that we recover the dispersion of the case with 3 bands, due to the indistinguishability of the 10 atoms in each layer.
The uniform intensity of the bands throughout the Brillouin zone is a direct consequence of our choice of $A_{l'l}$.

\begin{figure}[t]
  \centering
  \includegraphics[scale=0.75,clip=true,trim=1.3em 0.5em 0em 0em]{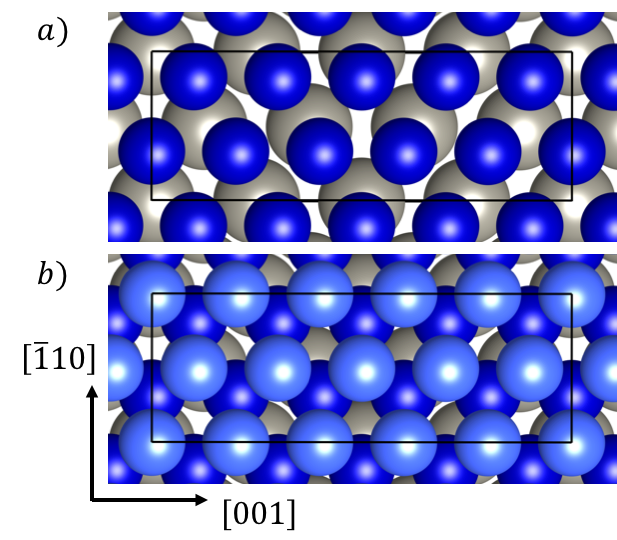} \vspace{-1em}
  \caption{\label{fig:structure} (a) Top view of a Co ML on W(110), in the $4 \times 1$ reconstruction.
  The dark blue spheres represent Co atoms, while the gray ones are W atoms.
  The crystallographic directions for bulk W are also indicated.
  There are five Co atoms covering four W atoms in the [001] direction.
  (b) 2ML Co on W(110) in hcp stacking.
  The Co layer at the interface is shown with dark blue spheres, while light blue spheres depicting the second Co layer.}
\end{figure}

\begin{figure}[t]
  \centering
  \includegraphics[scale=0.5,clip=true,trim=1.5em 0em 0em 0em]{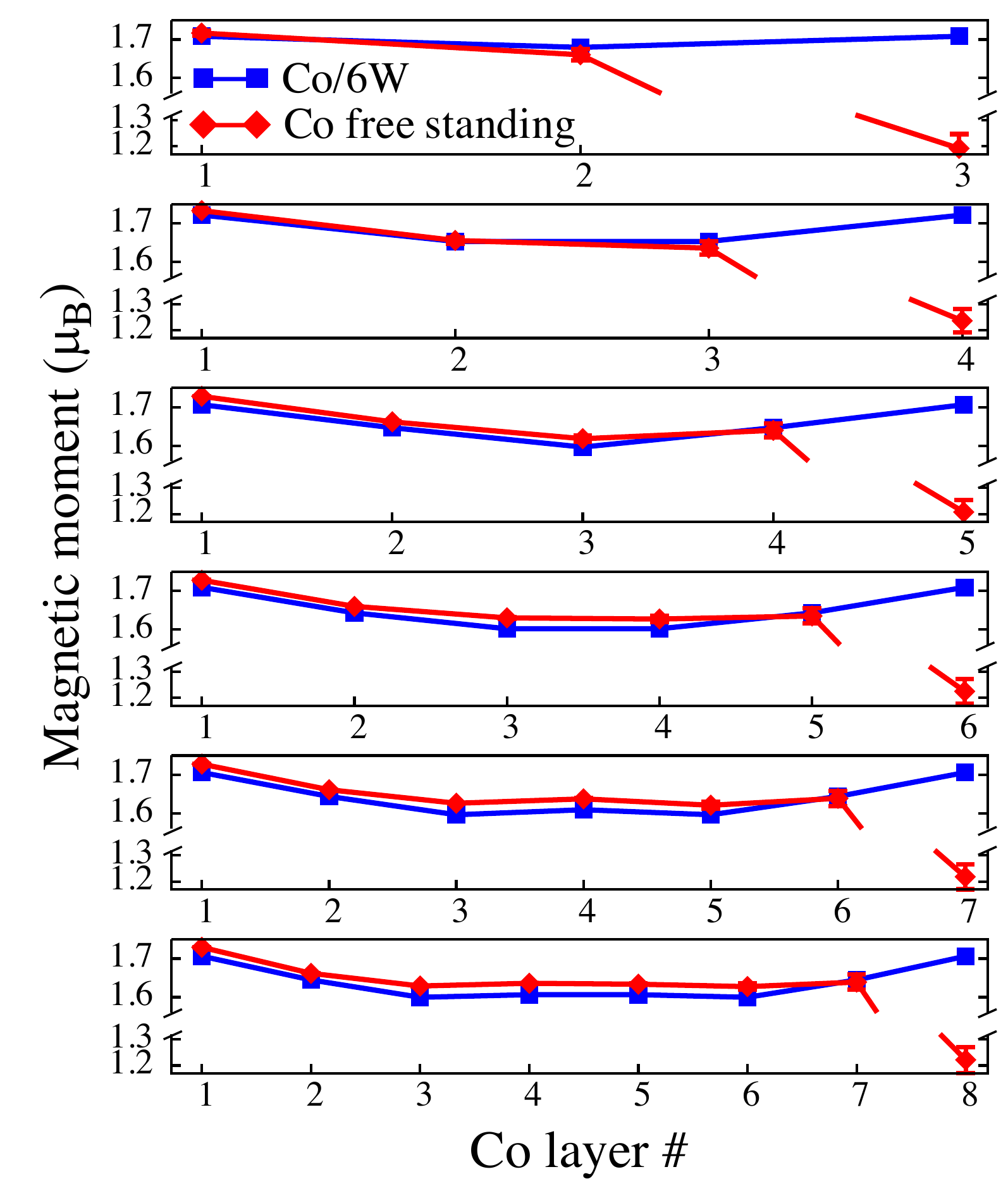} \vspace{-1em}
  \caption{\label{fig:magmom} Layer-resolved spin magnetic moments for free-standing and supported Co films.
  For the supported films with $n$ layers, Co layer 1 is the surface layer and Co $n$ is at the W(110) interface.
  The magnetic moments for the supported films are averaged over the ten Co atoms in each layer, with the error bar indicating the spread.
  }
\end{figure}

\begin{figure}[t]
  \centering
  \includegraphics[scale=0.4]{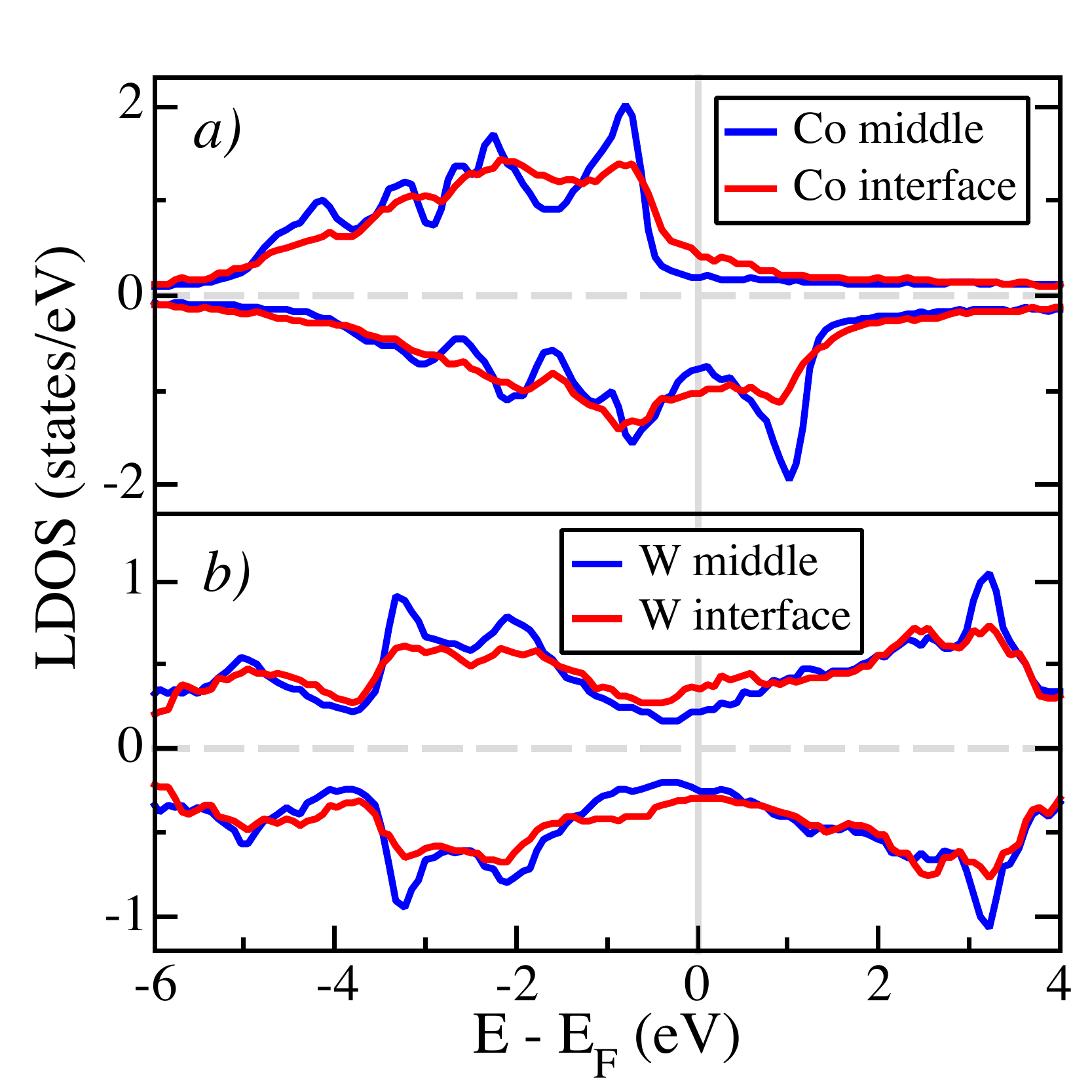} \vspace{-1em}
  \caption{LDOS for the 8Co/W(110) slab.
  The energy zero marks the Fermi energy.
  Positive values correspond to the majority spin LDOS, and negative ones to the minority spin LDOS.
  (a) Comparison of the average LDOS for the Co layers at the interface and in the middle of the Co film (bulk-like).
  (b) Comparison of the average LDOS for the W layers at the interface and in the middle of the W film (bulk-like).
  The smearing of bulk-like peaks and transfer of spectral weight to near the Fermi energy signal the strong Co--W hybridization at the interface.
  These changes lead to reduced magnetic moments for the Co layer at the interface, and also impact the magnetic exchange interactions.}
  \label{fig:dos}
\end{figure}

\begin{figure}[t]
  \centering
  \includegraphics[scale=0.4]{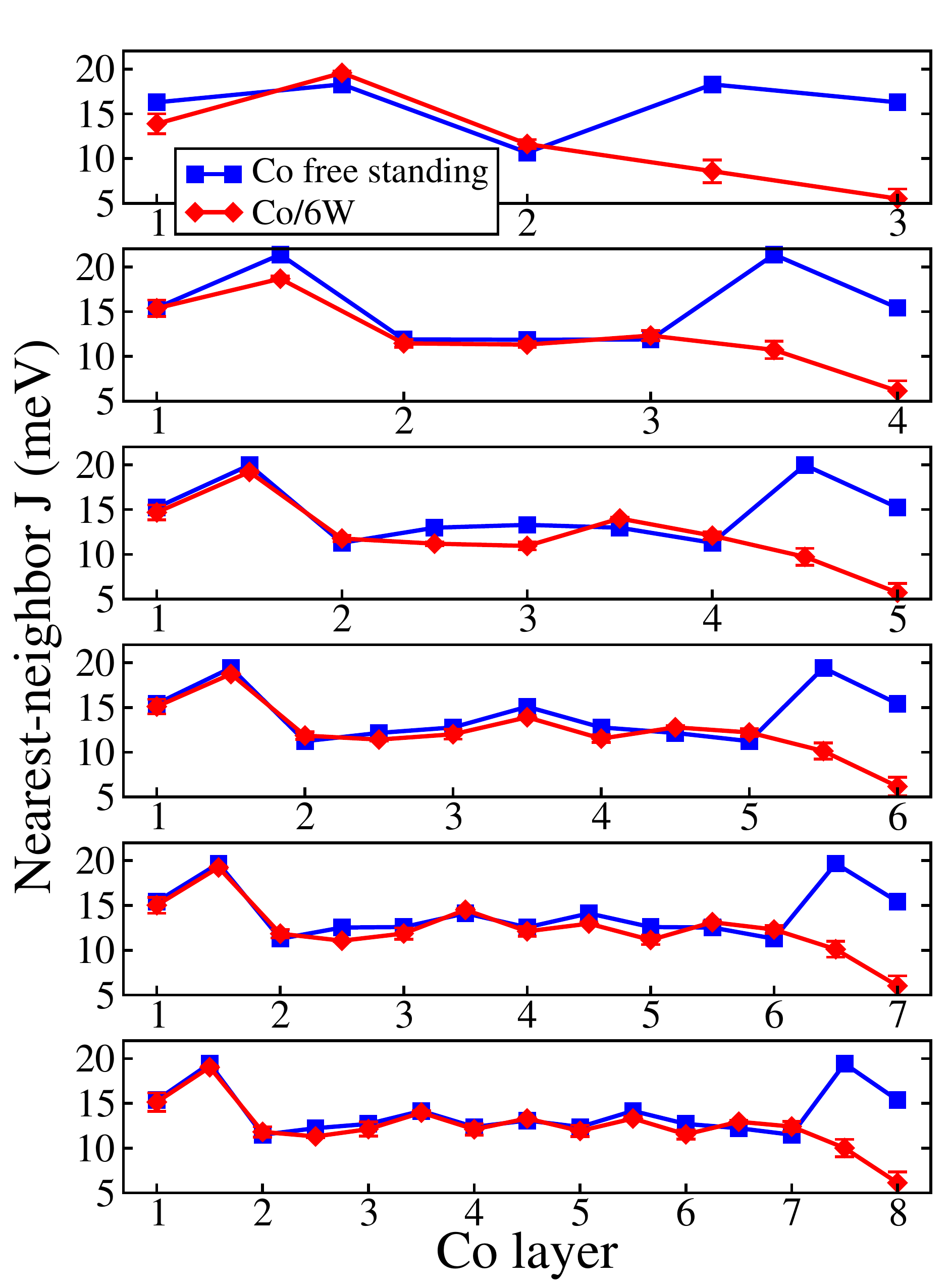} \vspace{-1em}
  \caption{Nearest-neighbour magnetic exchange parameter $J$ among Co atoms, for free-standing and supported Co films of different thickness.
  For Co/W(110), the interface layers are on the right-hand side. 
  The intralayer parameter for layer $n$ is labeled by the same integer, while the coupling between layers $n$ and $n+1$ is labeled by $n+1/2$. 
  For the supported films, the average $J$ is shown, with the spread given as an error bar.
  Due to the Co--W hybridization, the coupling strength decreases for the Co layer at the interface.}
 \label{fig:Jijnn}
\end{figure}

\begin{figure*}[t]
  \centering
  \includegraphics[width=\linewidth]{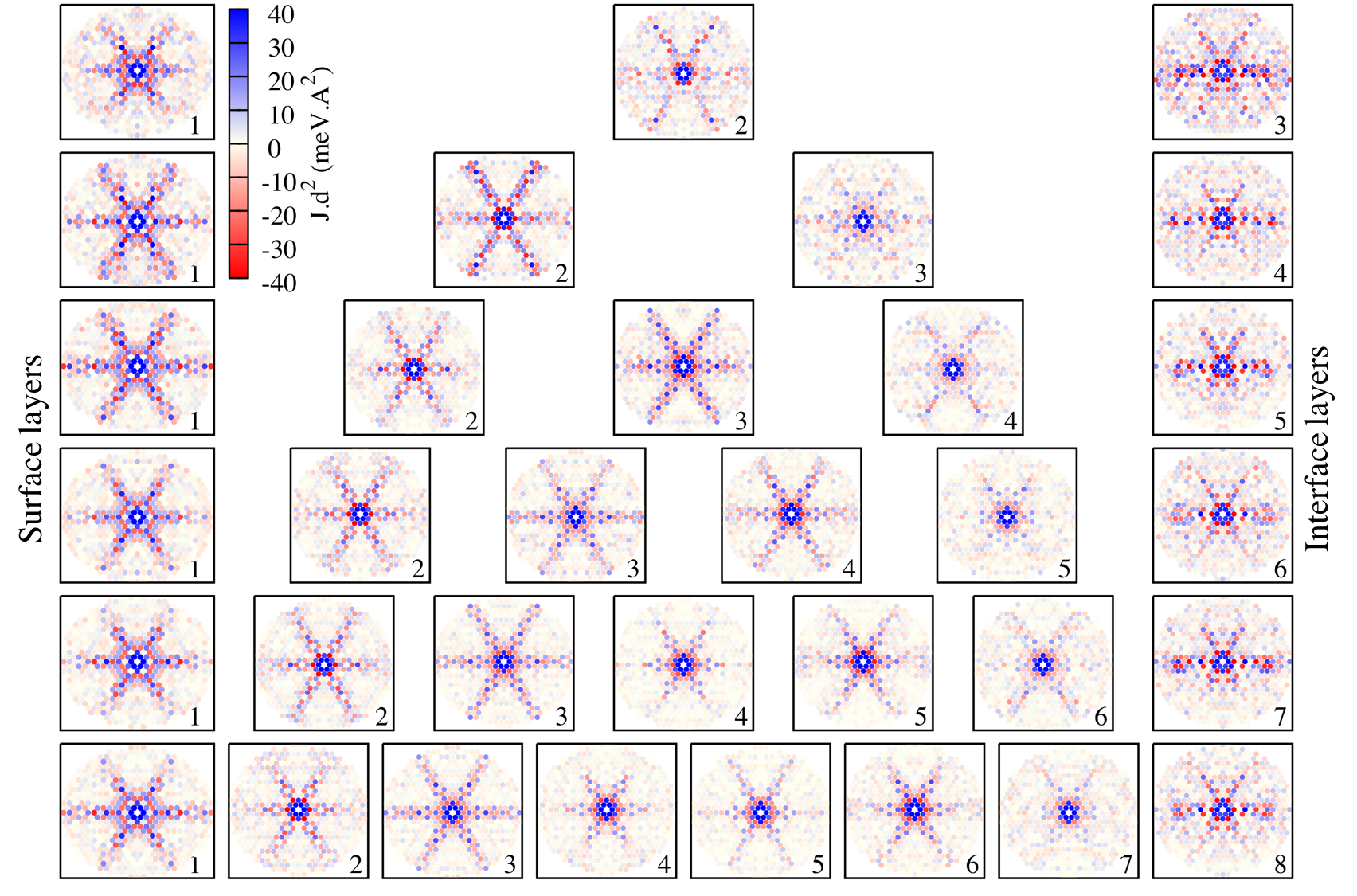} \vspace{-1em}
  \caption{Maps of the intralayer magnetic exchange interactions in real space, for 3-8ML Co on W(110).
  Each map shows the magnetic exchange interaction, $J_{ij}$, between the first Co atom in a given layer, $i$, and all other Co atoms in the same layer, $j$, up to a cutoff radius of \SI{30}{\angstrom}.
  The $J_{ij}$ are multiplied by $d^2$, where $d$ is the distance between the $i$ and $j$ atoms.
  The panels on the right-hand side correspond to those of Co layers at the vicinity of W.
  }
 \label{fig:Jijmap}
\end{figure*}

\begin{figure}[ht]
  \centering
  \includegraphics[width=0.48\textwidth, clip=true, trim=0.0em 2em 0em 6em]{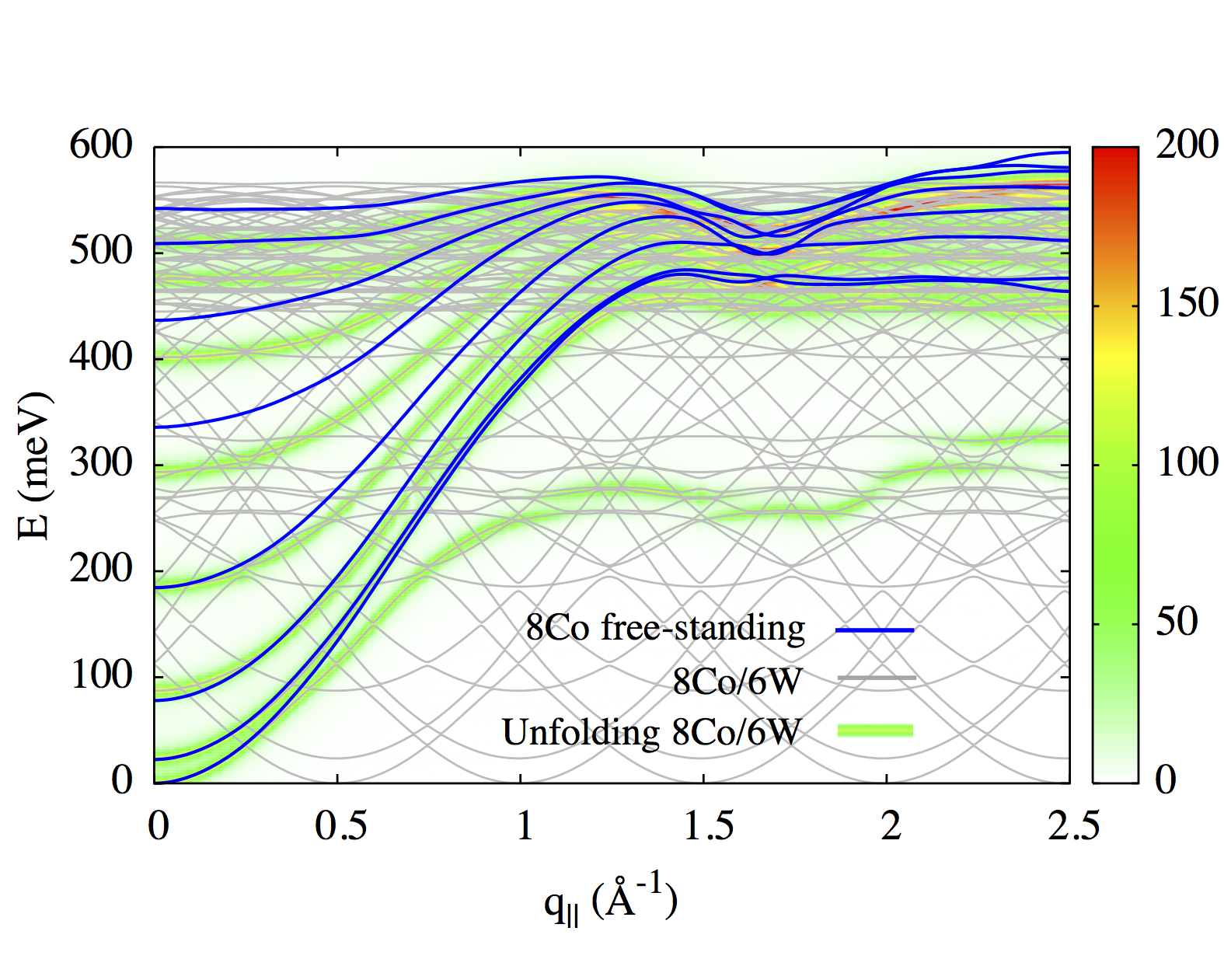}
  \vspace{-2em}
  \caption{\label{fig:freeVSsupported}
 Spin-wave dispersion for the free-standing and W-supported 8ML Co film (blue and gray lines, respectively).
 The color map corresponds to the unfolded dispersion for the supported films, Eq.~\eqref{eq:sfactor}, with a Lorentzian broadening of width $4\,\text{meV}$. The intensity of the color map is in arbitrary units.
  }
\end{figure}

\subsection{First-principles calculations}

The first-principles calculations are based on density-functional theory.
The atomic structure for Co/W(110) discussed in the next section was validated with Quantum Espresso~\cite{Giannozzi2009}, using the Projector Augmented Wave (PAW) method with a kinetic energy cutoff of $50\,\text{Ry}$, in the $\Gamma$-point approximation.
The magnetic moments and the magnetic exchange interactions are obtained with the Korringa-Kohn-Rostoker (KKR) Green-function method in the Local Spin Density Approximation (LSDA), and the atomic sphere approximation with full charge density (angular momentum cutoff $\ell_{\text{max}} = 3$)~\cite{Papanikolaou2002}.
We consider a slab geometry with open boundary conditions along the stacking direction, including two vacuum regions, each \SI{6}{\angstrom} thick.
The energy integration is performed in the upper complex energy plane,~\cite{Wildberger1995} with 30 points in a rectangular path and 5 Matsubara frequencies, for a temperature $T = \SI{500}{\kelvin}$.
The two-dimensional (2D) Brillouin zone integration was performed with a mesh of $30 \times 30$ and $5 \times 20$ $k$-points, for free-standing and supported films, respectively.
The magnetic exchange interactions are obtained from infinitesimal rotation of the magnetic moments as expressed in the Liechtenstein-Katsnelson-Antropov-Gubanov (LKAG) formula~\cite{Liechtenstein1987}.
For these calculations the number of Matsubara frequencies was increased to 10, with $T = \SI{100}{\kelvin}$, and the k-mesh was refined to $100 \times 100$ and $20 \times 80$, for free-standing and supported films, respectively.

\vspace{-1.em}

\section{Ground state properties} \label{sec:groundstate}
\subsection{Atomic structure}

We consider two kinds of systems: free-standing Co films comprising 3--8 monolayers (ML), with the bulk Co hcp structure, and Co films deposited on the W(110) surface with the same coverage range, but following a reconstructed hcp structure found experimentally~\cite{Knoppe1993,Ociepa1990, Spisak2005,Fritzsche1995,Pratzer2003}.
The free-standing Co films are used to identify which characteristics of the spin-wave dispersion arise from the reduced dimensionality and which can be attributed to the W(110) substrate.

hcp cobalt grows pseudomorphically on W(110), up to a coverage of 0.7 ML.
Beyond that, a reconstruction of the cobalt structure takes place due to the large lattice mismatch ($a_{\text{Co}} = \SI{2.51}{\angstrom}$, $a_{\text{W}} = \SI{3.16}{\angstrom}$).
The mismatch between the W(110) and the Co(0001) lattices is of 26\% in the W[001] direction and $3$\% in the [1$\overline{1}$0], and it is relieved by a $4 \times 1$ reconstruction, where five Co atoms cover four W atoms in the W[001] direction.
This corresponds to a stretching of the bulk Co(0001) lattice by 1\% along the W[001] and 3\% along the W[1$\overline{1}$0].
The resulting supercell contains 10 atoms in each Co layer and 8 in each W layer.
Possible in-plane modulations of the Co atomic positions and vertical relaxations have been considered in Ref.~\onlinecite{Spisak2005} and in our calculations.
They were found to have only a minor impact on the magnetic exchange interactions, so we adopted a simplified structural model.
Every Co atom in a given layer is at the same height, and sits on a slightly distorted hexagonal lattice, as shown in Fig.~\ref{fig:structure}(a); Fig.~\ref{fig:structure}(b) illustrates the hcp stacking of 2 ML Co on W(110).
The vertical interlayer distance was fixed at the bulk values similarly to the free standing Co films, $d_{\text{Co--Co}} = \SI{2.03}{\angstrom}$ and $d_{\text{W--W}} = \SI{2.23}{\angstrom}$, while at the interface $d_{\text{Co--W}} = \SI{2.13}{\angstrom}$.
The W(110) substrate is modelled using six W layers.
In total we have between 132 atoms (3Co ML) and 182 atoms (8Co ML) in our computational unit cell.

\subsection{Magnetic moments and electronic structure}

We begin the investigation of the impact of the interface with W(110) on the magnetic properties of Co thin films by analyzing some ground state properties.
Fig.~\ref{fig:magmom} compares the layer-resolved spin magnetic moments of free-standing Co films with the layer-averaged values for Co/W(110) films.
The magnetic moments for the supported Co films are very close to those of free-standing films of the same thickness, except for the Co layer at the interface with W(110).
There the magnetic moments are 30\% smaller, and there is some variability among the ten Co atoms comprising that layer, as indicated by the error bar in Fig.~\ref{fig:magmom}.
The interfacial W(110) layer acquires an average spin magnetic moment of $0.076\,\mu_{\text{B}}$, which is antiparallel to the Co magnetic moments and insensitive to the thickness of the Co film.

The explanation for the strong reduction of the magnetic moment of Co at the interface is found in the hybridization of the Co $d$-states with the W $d$-states, as seen in the layer-resolved density of states (LDOS), Fig.~\ref{fig:dos}.
Contrasting with the LDOS for bulk-like layers, there is an increase of spectral weight near the Fermi energy, which is responsible for the reduction of the spin magnetic moment of the Co interface layer.
A comparison with the electronic structure of free-standing films of the same thickness reveals that the LDOS for the other Co layers is only weakly disturbed by the presence of the W(110) interface; this also explains why the spin magnetic moments are very similar for both kinds of systems, except for the interface layer.

\subsection{Magnetic exchange interactions}

First we consider the nearest-neighbour interaction, see Fig.~\ref{fig:Jijnn}, where both the intralayer and the interlayer couplings are shown.
Significant changes are only apparent for the Co layer at the W(110) interface, for which we find reduced intralayer and interlayer couplings, comparing with the free-standing films.

The magnetic exchange interactions are fairly long-ranged, and they reflect the symmetry of the electronic states that give rise to them.
Fig.~\ref{fig:Jijmap} shows some representative cases: the intralayer magnetic exchange interaction between the first Co atom in a given layer and all the others in the same layer, up to a cutoff of \SI{30}{\angstrom}.
In this figure the value of $J$ is multiplied by $d^2$ ($d$ being the distance between atoms) to compensate for the decay with distance.
The oscillating sign changes with the distance lead to alternation between ferromagnetic and antiferromagnetic interactions.

For thicker films, the Co layers away from the W(110) interface reproduce the behavior of the freestanding Co films.
The slower decay along the six nearest-neighbour directions arises from the hexagonal shape of the Co $d$-bands in the Brillouin zone, near the Fermi energy.
This can be shown using simple arguments presented in Appendix~\ref{Apx:focusing}.
The presence of the interface modifies the long-range behavior of the magnetic interactions for the two Co layers next to it (see the panels on the right-hand side of Fig.~\ref{fig:Jijmap}).
This is in contrast with the change in the spin magnetic moments and nearest-neighbour magnetic exchange interactions, which are only significantly impacted for the Co layer in contact with W(110).
The symmetry of the pattern for slowly decaying interactions is also modified next to the interface, being reduced from hexagonal to twofold for the Co layer at the W(110) interface.

\section{Spin-wave dispersions}\label{sec:swdispersion}

\subsection{Free-standing versus supported films} 
\label{sub:free_standing_versus_supported_films}

Having characterized the ground state magnetic properties of the Co films, we can finally understand the properties of the spin-wave dispersions.
The spin-wave dispersions for the free-standing and supported Co thin films are calculated within the adiabatic approach described in Sec.~\ref{sec:theory}, initially for a thickness of 8ML, as shown in Fig.~\ref{fig:freeVSsupported}.
The blue lines are the results for the free-standing film, while the gray lines stand for the supported film.
The large number of spin-wave bands in the supported film is due to the lateral inhomogeneity arising from the surface reconstruction, and \textit{a priori} it is unclear how a comparison with the free-standing case can be made.
This is answered by the unfolding procedure summarized in {Eq.~\eqref{eq:sfactor}}, the result being shown as the background color map.
One can then focus on the unfolded dispersion of the supported film when comparing to the free-standing calculation.

\begin{figure}[t]
  \centering
  \includegraphics[width=0.45\textwidth,trim={1em 0 0 2em},clip=true]{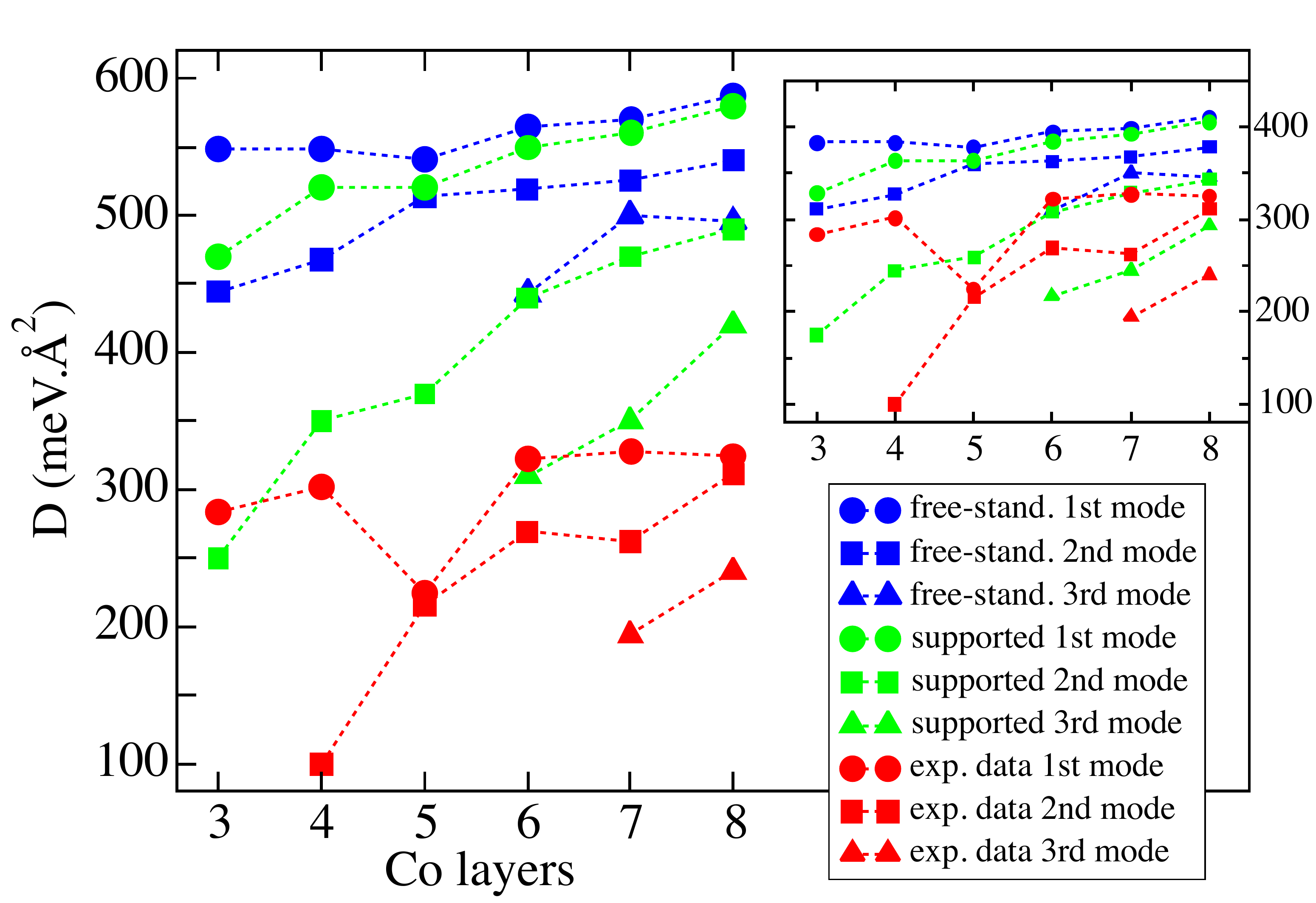} \vspace{-1em}
  \caption{Stiffness constants obtained from fitting the spin-wave branches of Fig.~\ref{fig:dispersions} to Eq.~\eqref{eq:quadratic}.
  Blue refers to the free-standing calculations while green stands for the supported case.
  Red indicates the stiffness of the experimental data\cite{Michel2015}.
  For all fits, only points with $q_\parallel < \SI{0.3}{\per\angstrom}$ were considered.
  Circles correspond to the first (acoustic) mode, squares to the second mode, and triangles to the third one.
  The inset presents the same data as above with free-standing and supported results rescaled down by 30\%, as will be discussed in Sec.~\ref{sub:theory_versus_experimental_results}.
  The supported films capture the experimental trends better than the free-standing ones. 
  }
 \label{fig:stiffness}
\end{figure}

There are two main points of interest when comparing the free-standing and supported calculations.
First we consider the spin-wave energies at $q_\parallel=0$.
For example, the second spin-wave branch (first optical mode) for the 3ML free-standing film is slightly higher in energy than for the 3ML W-supported film, while for the 4 and 5ML thicknesses the ordering is reversed.
However, these energy differences decrease for thicker slabs (which is also true for other modes). 
These energy gaps between the different modes at $q_\parallel=0$ are mainly determined by the interlayer exchange coupling, which is modified only near the Co/W interface.
For thicker films, the contribution from the interface becomes less important, and both free-standing and supported films should become similar.
Also, the higher-energy modes are more strongly affected by the substrate.
Returning to the 8Co/W film of Fig.~\ref{fig:freeVSsupported}, we observe that the first four modes are very close to the corresponding free-standing ones in the small-$q$ region, while the higher modes still differ.

The second point of interest is the stiffness of the spin-wave branches, which indicates how strongly the spin-wave energy increases with the wavevector.
We find that, for all modes and all thicknesses, the stiffness is larger for the free-standing films than for the supported ones.
Therefore, the substrate softens the spin-waves.
This is more pronounced for the first mode (also known as the acoustic mode), where the result of the two calculations spread apart for wavevectors larger than about \SIrange{0.3}{0.4}{\per\angstrom}, as can be seen in Fig.~\ref{fig:freeVSsupported}.

In the small wavevector regime, the spin-wave dispersions are quadratic, 
\begin{equation}\label{eq:quadratic}
  E_n(q_\parallel) \approx E_n(0) + D_n q_\parallel^2 \qquad ,
\end{equation}
where $D_n$ is the stiffness constant of the $n$-th mode.
A full comparison between the stiffnesses of different modes in different thicknesses for both free-standing and supported films is shown in Fig.~\ref{fig:stiffness}.
In addition, Fig.~\ref{fig:stiffness} also shows the experimental stiffnesses extracted from the dispersion published in Ref.~\onlinecite{Michel2015}.
We defer the comparison between theory and experiment to the next section, where the experimental data is also plotted in {Fig.~\ref{fig:dispersions}}.
The quadratic fit to Eq.~\eqref{eq:quadratic} was applied to the dispersion curves in the range $q_\parallel \in [0.0,0.3]\,\si{\per\angstrom}$.
As pointed out before, one can observe that systematically the free-standing films have higher stiffness, with the differences to the supported films being larger for higher modes and thinner films.

\begin{figure}[ht]
  \centering
  \includegraphics[width=0.48\textwidth,trim={1em 0 0 0},clip=true]{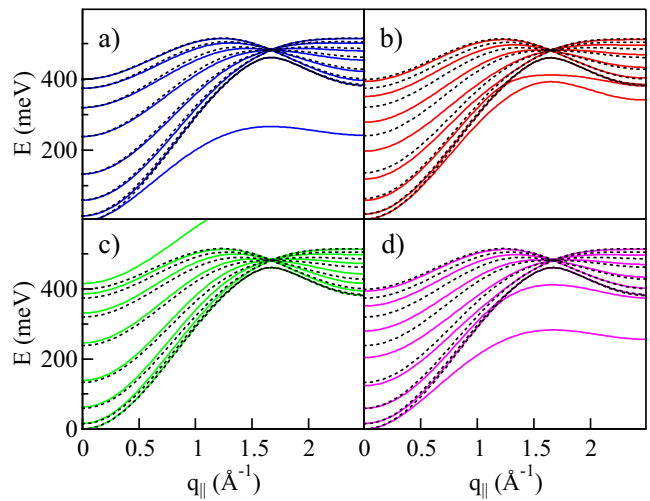} \vspace{-1em}
  \caption{Dispersion curves for a nearest-neighbor Heisenberg model of an 8ML Co film, based on the parameters given in Figs.~\ref{fig:magmom} and \ref{fig:Jijnn} for the free-standing films.
  The black-dashed lines in all panels represent the result obtained with unmodified parameters, while the solid lines show the dispersion upon the following changes:
  (a) The intralayer coupling of the last Co layer is reduced by $60\%$.
  (b) The interlayer coupling between the interface Co layer and its adjacent layer is reduced by $50\%$.
  (c) The magnetic moment of the interface Co layer is reduced by $30\%$.
  (d) The effect of combining all the changes in the parameters.}
 \label{fig:dispersion_model}
\end{figure}

To unravel the impact of the substrate on the spin-wave dispersions, we discuss a simplified magnetic interaction model, using only nearest-neighbor couplings.
We have learned in Sec.~\ref{sec:groundstate} that the tungsten substrate mainly decreases the following: (a) the intralayer coupling of the Co interface layer (Fig.~\ref{fig:Jijnn}); (b) the interlayer coupling between the Co interface layer and the adjacent Co layer (Fig.~\ref{fig:Jijnn}); and (c) the magnetic moment of the Co interface layer (Fig.~\ref{fig:magmom}).
To establish the qualitative impact of each of these factors on the spin-wave dispersions, we parametrized a nearest-neighbor model for a 8ML film with the data of Figs.~\ref{fig:magmom} and \ref{fig:Jijnn}, which pertain to the free-standing case.
The resulting dispersions are shown with black-dashed lines in all panels of Fig.~\ref{fig:dispersion_model}, and they will serve as reference.

In Fig.~\ref{fig:dispersion_model}(a), we decreased by $60\%$ the intralayer exchange coupling of the last Co layer (ratio taken from Fig.~\ref{fig:Jijnn}).
Comparing with the reference model (black-dashed), the main difference is the strong reduction of the acoustic mode stiffness, and the lowering of its energy throughout the Brillouin zone.
Fig.~\ref{fig:dispersion_model}(b) shows the impact of decreasing by $50\%$ only the interlayer coupling between the last Co layer and the adjacent one.
Almost all spin-wave branches are modified, and their energy lowered, but the overall bandwidth is mostly preserved.
The acoustic mode is lowered only next to the border of the Brillouin zone (the crossing point of the higher branches).
Fig.~\ref{fig:dispersion_model}(c) reveals that reducing only the magnetic moment of the last Co layer by $30\%$ (Fig.~\ref{fig:magmom}) leads to very small deviation from the reference model, except for the highest branch, which gets pushed higher in energy.
Lastly, Fig.~\ref{fig:dispersion_model}(d) shows the result of combining all three modifications.
Most of the characteristics of the full calculations of Fig.~\ref{fig:freeVSsupported} are present: changes in $E_n(0)$, the reduction of the stiffness of the modes, and the lowering of the acoustic mode.


\subsection{Theoretical vs experimental dispersion} 
\label{sub:theory_versus_experimental_results}

\begin{figure*}[ht]
  \centering
  \includegraphics[width=0.9\textwidth, clip=true,trim=0.0em 0em 0em 0em]{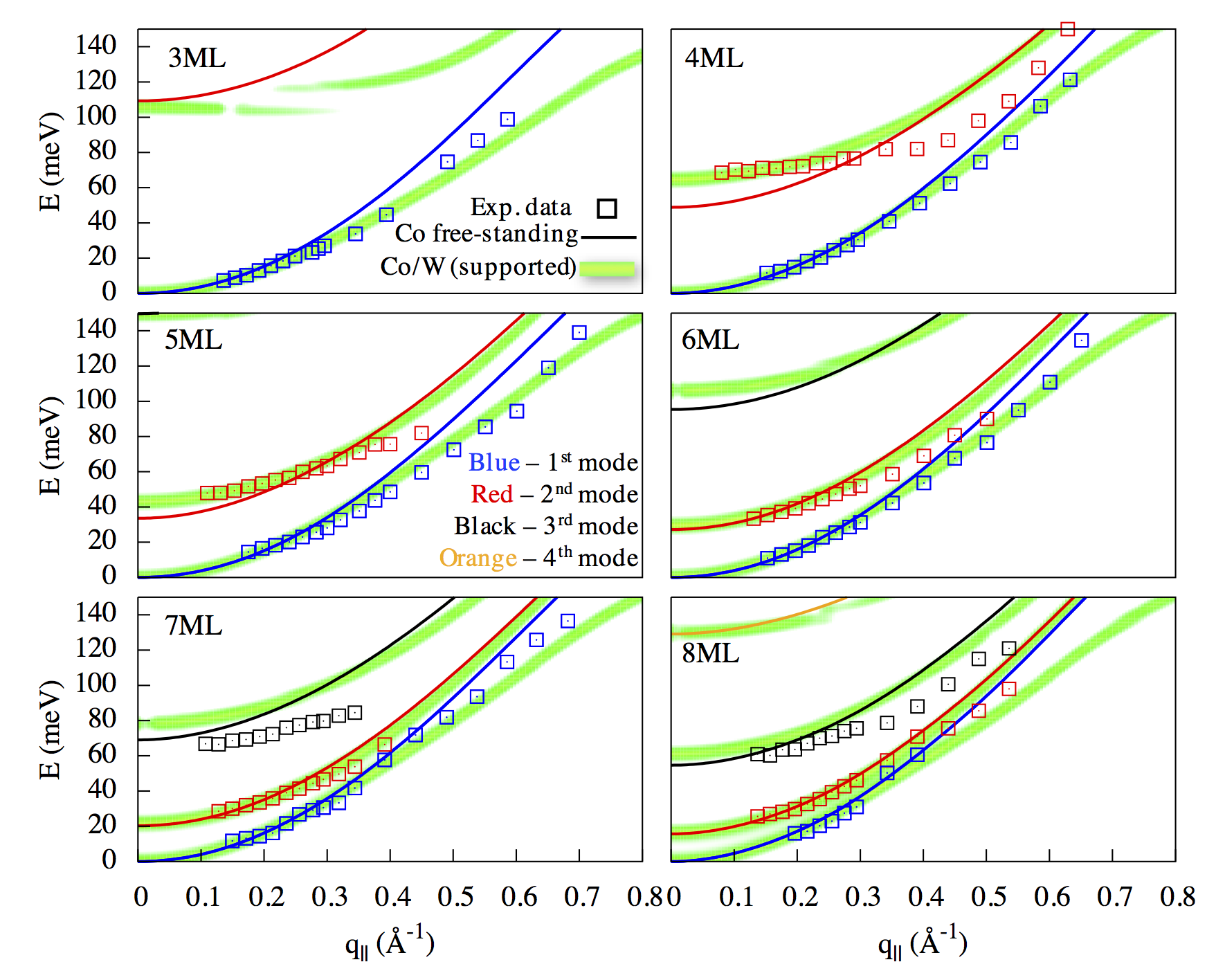} \vspace{-1em}
  \caption{\label{fig:dispersions}
  Comparison of calculated (lines) and experimentally measured spin-wave dispersions (squares, from Ref.~\onlinecite{Michel2015}) for several thicknesses.
  The thin lines are the spin-wave branches obtained for the free-standing films, while the thick green-yellow lines (actually a color map) correspond to the unfolded  dispersion for the W-supported films, Eq.~\eqref{eq:sfactor}.
  In the unfolding scheme, a Lorentzian broadening of width $4\,\text{meV}$ was considered.
  The magnetic exchange coupling has been uniformly rescaled down by 30\si{\percent}.
  }
\end{figure*}

We have seen that the theoretical stiffnesses are systematically higher than the experimental ones~\cite{Michel2015}.
Fig.~{\ref{fig:stiffness}} shows that the stiffnesses of the acoustic mode range from 450 to {\SI{600}{\milli\electronvolt\angstrom\squared}} for both free-standing and supported calculations, while the experimental ones range from 200 to {\SI{300}{\milli\electronvolt\angstrom\squared}}.
One possible reason for the discrepancy is the sensitivity of the fit to the available experimental data.
On the one hand, the number of experimental data points in this range is rather small, and the spin-wave energies cannot be experimentally determined for $q_\parallel \rightarrow 0$ due to limitations of the EELS technique~{\cite{Michel2015}}.
On the other hand, including experimental data at higher $q_\parallel$ also increases the uncertainty in the fitting procedure, due to the difficulty in extracting the spin-wave energies from broad experimental peaks~{\cite{Michel2015}}.

It is also known that DFT in the LSDA overestimates the exchange splitting of metallic ferromagnets, such as Co and Ni, and consequently their magnetic moments and magnetic exchange coupling.
In the work of M\"uller \textit{et al.}~\cite{Muller2016}, it is reported that LSDA calculation for bulk fcc cobalt leads to an exchange splitting 30\% higher at $\Gamma_{25}^\prime$ (even 55\% at $\Gamma_{12}$) with respect to the experimental value.
The same work points to a way for an improved description of the electronic structure, based on many-body perturbation theory.
However, such methods are already computationally very demanding for bulk systems containing just a few atoms in the unit cell, which makes them unfeasible for the structurally complex thin films we considered.

Ref.~{\onlinecite{Muller2016}} also pointed out that an alternative is to rescale the exchange splitting self-consistently in the LSDA calculation, $B_{\text{xc}} \rightarrow \alpha\,B_{\text{xc}}$, which then renormalizes the magnetic parameters of the Heisenberg model computed from first-principles.
Unfortunately, the magnitude of the rescaling is unknown \textit{a priori}.
The magnetic interactions are affected in a nonlinear way by $\alpha$, as we verified in our calculations.
We note that an empirical reduction of $J(\VEC{q})$ by 15\% was already explored in Ref.~{\onlinecite{Rajeswari2014}}, to bring theoretical and experimental results for fcc Co/Cu(001) into agreement.
For free-standing hcp cobalt films, we observed that reducing the exchange splitting up to 20\% ($\alpha=0.8$) has an overall effect of rescaling the exchange interactions, but by a different factor, $J(\VEC{q}) \rightarrow \beta\,J(\VEC{q})$.
For an 8ML free-standing Co film, a reduction of the magnetic interactions by 30\% ($\beta=0.7$) is approximately obtained from a rescaling of the exchange splitting in the 10-20\% range ($\alpha \in [0.8,0.9]$).

Fig.~{\ref{fig:dispersions}} shows the theoretical results with the 30\% reduction of $J$, together with the experimental data of Ref.~{\onlinecite{Michel2015}}.
We obtain very good agreement with the experimental results, in particular for the supported films.
A single rescaling parameter is enough to describe well both the energies of the standing modes ($q_\parallel=0$) and the stiffnesses, for all film thicknesses and modes (only small deviations remain for the third mode of 7 and 8ML films).
The inset in Fig.~\ref{fig:stiffness}, comparing the experimental stiffnesses with the rescaled theoretical ones, highlights that our results for the supported films capture much better the trends in the experimental data.
Such a simultaneous match can not be achieved with the free-standing films, even by changing $\beta$ arbitrarily.
For example, if we adjust $\beta$ to obtain good agreement for the optical mode energies at $q_\parallel=0$ of the 4-5ML films, then the computed dispersions become much stiffer than the experimental ones; the optical mode energies at $q_\parallel=0$ for 6-8ML films that were already matching well would then go off.
As explained in the previous sections, the Co-W hybridization at the interface endows the supported film dispersions with the right features, $q_\parallel=0$ energies and stiffnesses, reproducing the characteristics of the experimental data.
For reference, a direct comparison of the theoretical results without rescaling with the experimental measurements can be found in Appendix B.


\section{Discussion and conclusions}\label{sec:conc}

The agreement between theoretical calculations and experimental measurements, shown in Fig.~\ref{fig:dispersions}, required a rescaling of the magnetic interactions strength, attributed to the expected overestimation of the spin splitting in the first-principles calculations.
We have explored other possibilities for the discrepancy between theory and experiment.
One might wonder if the failure lies with the adiabatic approach for the calculation of the spin-wave excitations.
Ref.~{\onlinecite{Buczek2010}} performed calculations for an Fe ML on W(110), comparing the results of the adiabatic approach to those including the coupling to the Stoner continuum, and found no substantial differences.
However, this can be system-dependent.
Recently, another possible explanation was put forward: finite-temperature softening of the spin-wave dispersion, as seen by calculating the dynamical structure factor for Fe overlayers on Ir(001)~\cite{Rodrigues2016}.
The idea is that temperature leads to a finite canting angle of the neighboring magnetic moments, which can reduce the strength of the magnetic exchange interactions~{\cite{Szilva2013}}.
However temperature can only play a role if the Curie temperature is close to the experimental temperature, which does not seem to be the case for Co/W(110), judging from the strength of the magnetic interactions.
In short, the fault seems to lie with the LSDA approximation, and a computationally efficient first-principles correction to the spin splitting remains to be found.

We demonstrate in our work that the interface matters in determining the dispersion of the spin-waves of the entire magnetic thin film.
Our first-principles calculations have provided an extensive theoretical characterization of the impact of the tungsten substrate on the spin-waves of the cobalt ultrathin films.
We found that only the Co layer directly at the interface with W is strongly affected, leading to a reduced spin moment, and weakened intralayer and interlayer magnetic exchange interactions.
The qualitative differences between the spin-wave dispersions of free-standing and W-supported films are well explained by a simple nearest-neighbor Heisenberg model, which takes into account the changes in the magnetic properties of the Co layer at the interface.
Taking into account the likely overestimated spin splitting of Co in the first-principles calculations, we found that good agreement with available EELS measurements could be reached for a realistic reduction of the strength of the magnetic exchange interactions.

\begin{acknowledgments}
We thank A.~T.~Costa, F.~S.~M.~Guimar\~{a}es and J.~Azpiroz for fruitful discussions.
We acknowledge E. Michel and H. Ibach for providing and discussing their experimental data, and the latter for carefully reading our manuscript.
This work is supported by the Brazilian funding agency CAPES under Project No. 13703/13-7, the Helmholtz-Gemeinschaft–Young Investigator Group Programme VH-NG-717 (Functional Nanoscale Structure and Probe Simulation Laboratory–Funsilab), and the European Research Council (ERC) under the European Union’s Horizon 2020 research and innovation programme (ERC-consolidator Grant No. 681405–DYNASORE).
\end{acknowledgments}


\appendix

\section{}\label{Apx:focusing}

\begin{figure}[tb]
  \centering
  \includegraphics[width=0.48\textwidth,trim={1.8em 0 0 0},clip=true]{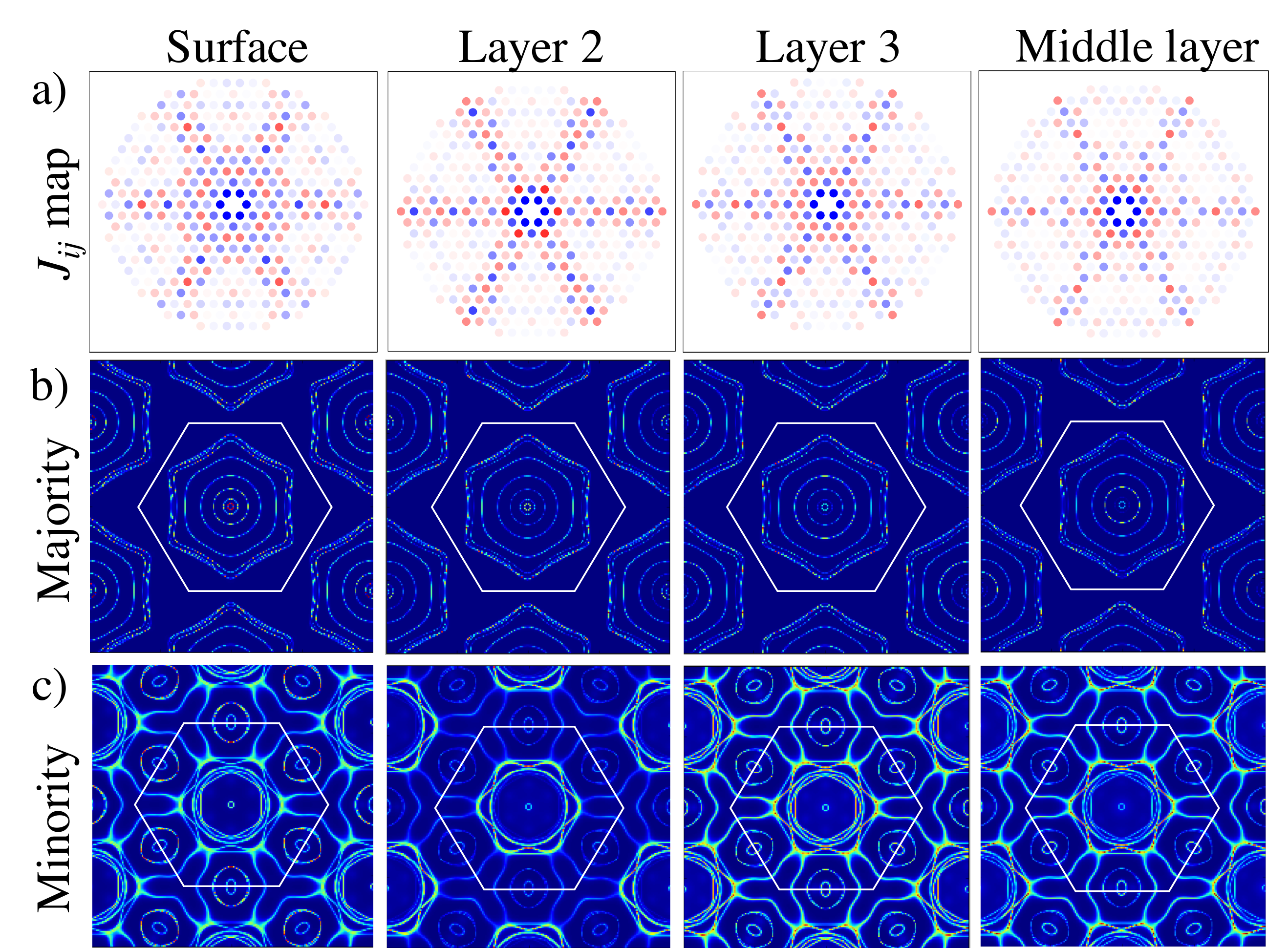}
  \caption{\label{fig:fermisurf_mapjs_slab}
  Panel (a) shows the intralayer $J_{ij}$ maps for the first layers of a 8ML Co free-standing slab.
  Panels (b) and (c) present the correspondent layer-resolved Fermi surface contours for the majority and minority spin channels, respectively.
  The $J_{ij}$ are multiplied by $d^2$, where $d$ is the distance between the $i$ and $j$ atoms.
  }
\end{figure}

The LKAG formula~\cite{Liechtenstein1987} provides the connection between the electronic structure and the magnetic exchange interactions.
In this appendix, we explain the anisotropic spatial dependence of the $J_{ij}$'s seen in Fig.~\ref{fig:Jijmap}.
We first consider the shape of the layer-resolved Fermi surface contours for the free-standing Co 8ML slab. 
The results are shown in Fig.~\ref{fig:fermisurf_mapjs_slab} for the first four layers (the other four are equivalent due to the mirror symmetry of the free-standing film), together with the respective intralayer maps of the magnetic interactions.
The Fermi surface of the majority-spin channel features circular contours in the center of the Brillouin zone and hexagon-like ones away from the center.
These hexagon-like bands have flat regions, which enhances the group velocity of the occupied electronic states, mediating an enhanced magnetic interaction for pairs of atoms aligned with their group velocity.

To corroborate our interpretation, we make use of a simple tight-binding model with a single orbital per atom forming a 1ML hexagonal lattice.
The Hamiltonian reads
\begin{equation}\label{eq:Hamiltonian_tbmodel}
\HH=t\sum_{\langle ij \rangle \sigma} a_{i\sigma}^\dagger a_{j\sigma}  +
  U\sum_{ i }  (n_{i\uparrow} - n_{i\downarrow}) \quad ,
\end{equation}
where $t$ is the hopping parameter that connects nearest-neighbour atoms, and $U$ creates the spin splitting of the two bands.
The  operator $a_{i\sigma}~(a_{i\sigma}^\dagger)$ annihilates (creates) an electron with spin $\sigma$ on site $i$, and $n_{i \sigma} = a_{i\sigma}^\dagger a_{i\sigma}$ is the number operator.
Using the translational symmetry of the system, the Hamiltonian can be transformed to
\begin{equation}
  \HH = \sum_{\mathbf k\sigma} H_{\mathbf k \sigma} = \sum_{\mathbf k \sigma} \left(t_\mathbf k + U_\sigma\right) a_{\mathbf k\sigma}^\dagger a_{\mathbf k\sigma}
\end{equation}
where $U_\sigma = +U , -U$ for $\sigma = \uparrow, \downarrow$, respectively. And 
\begin{equation}
 t_\mathbf k = t  \sum_{\langle i,j \rangle } e^{\mathrm{i}  \mathbf k \cdot \mathbf R_{ij}} \quad \text{and} \quad 
 a_\mathbf k = \frac{1}{\sqrt N} \sum_{ j } e^{\mathrm{i}  \mathbf k \cdot \mathbf R_{j}} a_{j} \quad .
\end{equation}

From this model one can easily calculate the magnetic exchange coupling via the LKAG formula:
\begin{equation}
  J_{ij}= \frac{U^2}{\pi} \int^{E_F} \hspace{-1em}  \Tr [ G_{ij\uparrow}(E) G_{ij\downarrow}(E)]\;\mathrm{d}E \quad ,
\end{equation} \label{eq:Liechtenstein}
where the real space Green-function is given as
\begin{equation}
  G_{ij\sigma}(E) = \frac{1}{\Omega_{BZ}} \int \hspace{-0.2em} e^{-i \mathbf k \cdot \mathbf R_{ij}} G_{\mathbf k\sigma}(E) \;\mathrm{d}\mathbf k \quad ,
\end{equation}
and
\begin{equation}
  G_{\mathbf k\sigma}(E) = \frac{1}{E - E_0 - H_{\mathbf k\sigma} + \mathrm{i} \eta} \quad ,
\end{equation}
with $\eta \rightarrow 0$.

\begin{figure}[h!]
  \centering
  \includegraphics[width=0.4\textwidth,trim={0em 0 0 0},clip=true]{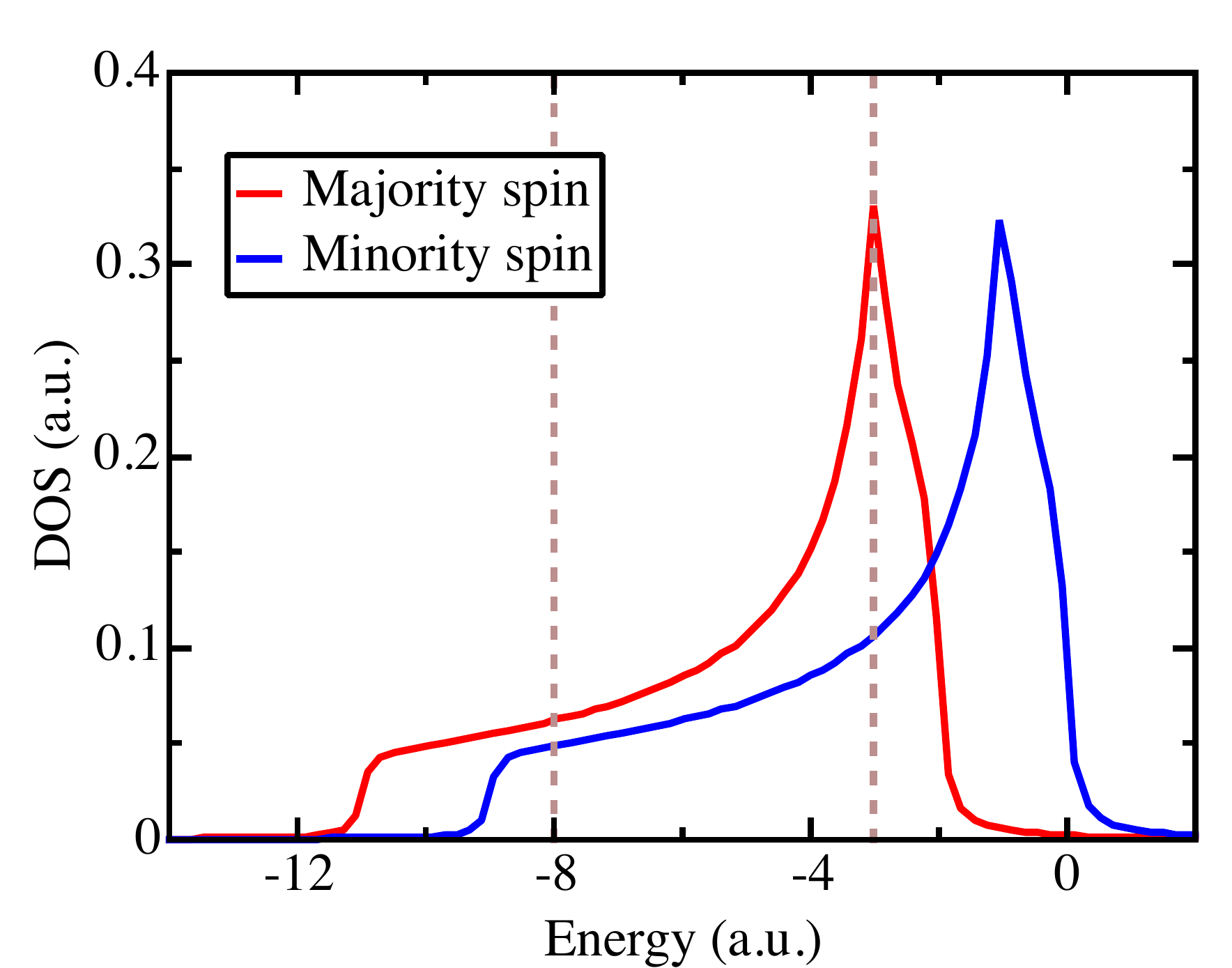}
  \caption{\label{fig:dosmodel}
  Total density of states of the two-band model given in Eq.~\eqref{eq:Hamiltonian_tbmodel}. 
  The dashed lines mark a Fermi energy near the bottom of the two bands (left), and another at the Van Hove singularity (right). The correspondent Fermi surface is almost isotropic in the first case, and very anisotropic in the other; see Fig.~\ref{fig:fermisurface}.  
  $t=-1$, $U=1$, $E_0=-4$, and $\eta=0.1$ a.u.
  }
\end{figure}

\begin{figure}[ht]
  \centering
  \includegraphics[width=0.5\textwidth,trim={0em 0 0 0},clip=true]{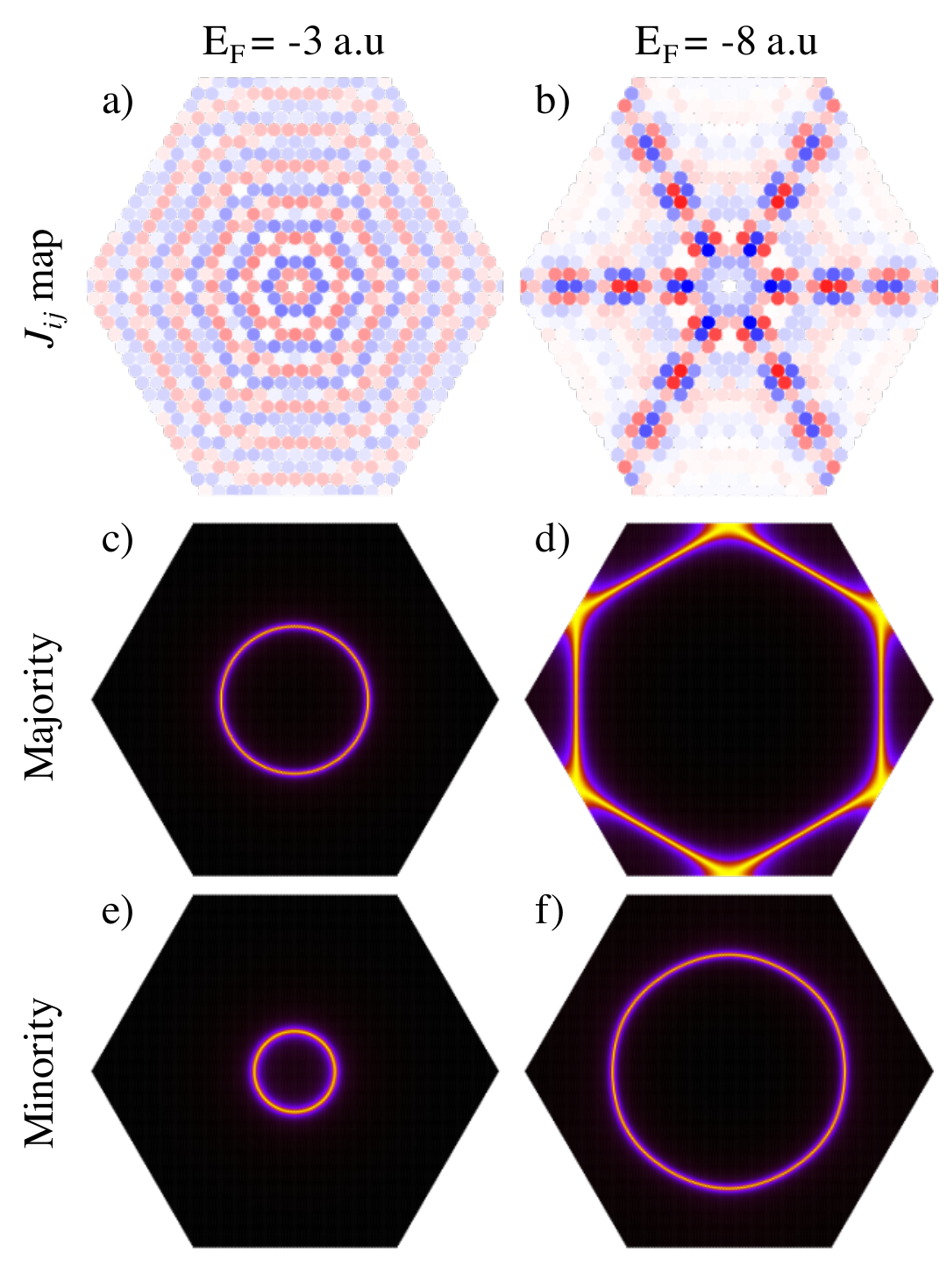}
  \caption{\label{fig:fermisurface}
  $J_{ij}$ maps for two different Fermi energies (a) $E_F=-8$~a.u. and (b) $E_F=-3$~a.u.
  Panels (c) and (e) are the majority and minority Fermi surface contours for $E_F=-3$~a.u., respectively, and panels (d) and (f) are the majority and minority Fermi surface contours for $E_F=-8$~a.u. 
  The $J_{ij}$ are multiplied by $d^2$, where $d$ is the distance between the $i$ and $j$ atoms.
  $t=-1$, $U=1$, $E_0=-4$, and $\eta=0.1$ a.u.
  }
\end{figure}

The DOS of this model is shown in Fig.~\ref{fig:dosmodel}.
Fig.~\ref{fig:fermisurface} displays the Fermi surface contours for Fermi energies marked in Fig.~\ref{fig:dosmodel}.
In the left column (panels (a), (c) and (e)), the Fermi energy is set near the bottom of the two bands, where the energy band dispersion is almost isotropic, see panels (c) and (e).
In the right column (panels (b), (d) and (f)), the Fermi energy is chosen to match the Van Hove singularity of the majority spin band, arising from the hexagonal shape of the energy dispersion.
The real-space map of the magnetic exchange interactions for both cases is shown in panels (a) and (b).
Panel (a) shows a very isotropic map, mainly marked by the periodic radial oscillation associated with Fridel oscillation.
Panel (b) shows the impact of the hexagonal shape of the energy bands near the Fermi energy, featuring a sixfold-symmetric focusing pattern.
We thus have illustrated our proposition that the anisotropy of the magnetic exchange interactions in real space is a direct consequence of the anisotropy of the electronic energy bands in reciprocal space that mediate the interactions (see similar effects obtained with adatoms in Refs.~\cite{Pruser2014,Bouhassoune2014,Weismann2009_2,Lounis2011}).


\begin{figure*}[ht]
  \centering
  \includegraphics[width=0.9\textwidth]{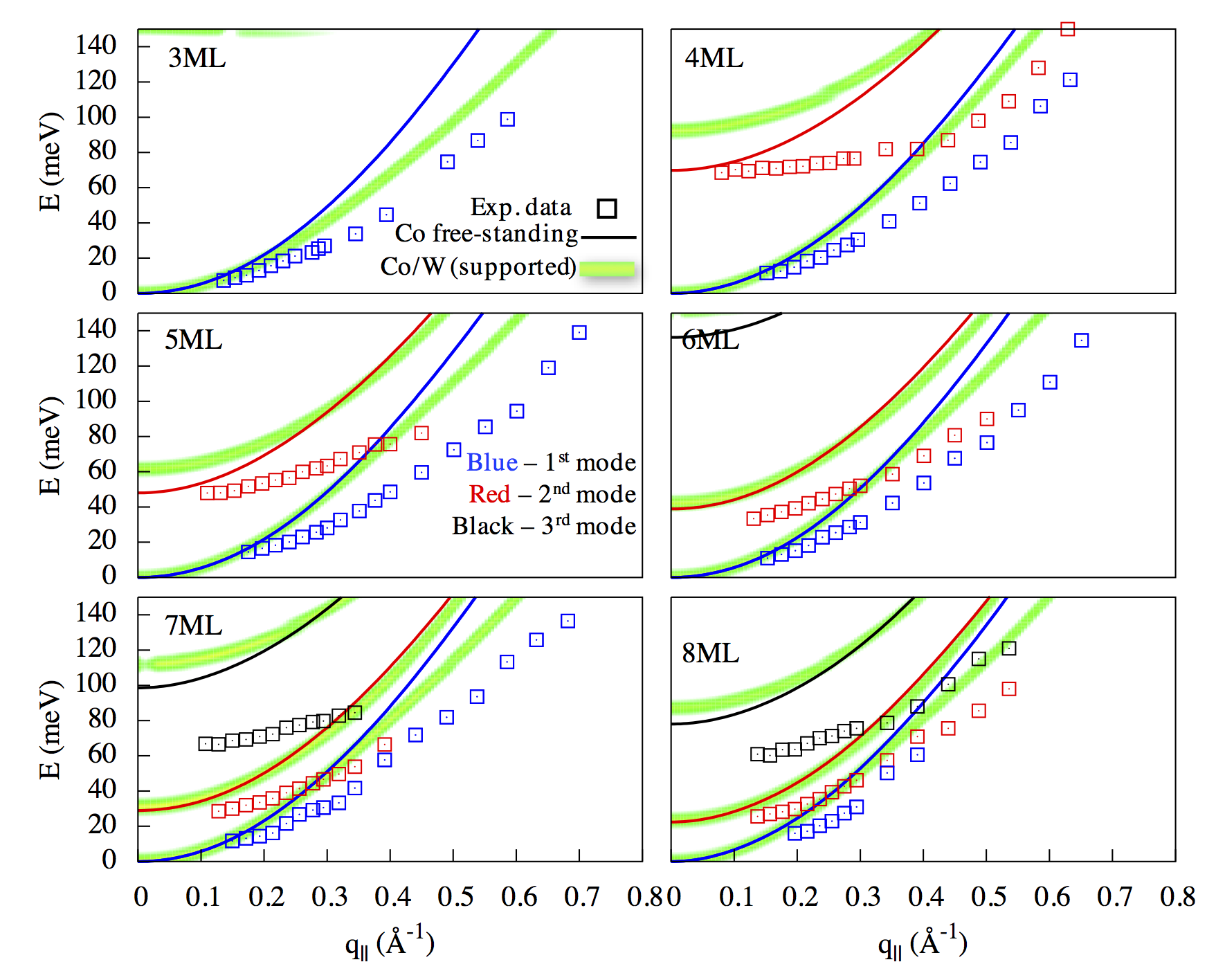} \vspace{-1em}
  \caption{\label{fig:dispersions2}
  Comparison of calculated (lines) and experimentally measured spin-wave dispersions (squares, from Ref.~\onlinecite{Michel2015}) for several thicknesses.
  The thin lines are the spin-wave branches obtained for the free-standing films, while the thick green-yellow lines (actually a color map) correspond to the unfolded  dispersion for the W-supported films, Eq.~\eqref{eq:sfactor}.
  In the unfolding scheme, a Lorentzian broadening of width $4\,\text{meV}$ was considered.
  }
\end{figure*}

\section{}\label{Apx:dispersion}

In Fig.~\ref{fig:dispersions} we have shown that a comparison between the theoretical results with the exchange coupling reduced by 30\% and the experimental data of Ref.~\onlinecite{Michel2015}, led to very good agreement, especially for the calculation of cobalt deposited on tungsten.
Here, we also present a direct comparison using unscaled parameters obtained by first principles; see Fig.~\ref{fig:dispersions2}.
It is clear that the spin-wave stiffnesses and $q_\parallel=0$ energies are overestimated in the theoretical results, which led us to explore possible explanations for this disagreement, as described in the main text.

\bibliography{bibliography}

\end{document}